\documentclass[aps,pra,reprint,groupedaddress,amsmath,longbibliography]{revtex4-1}

\usepackage{graphicx}
\usepackage{dcolumn}
\usepackage{amsmath}
\usepackage{amssymb}
\usepackage{lipsum}

\begin{document}


\title{Magic density in a self-rephasing ensemble of trapped ultracold atoms}




\author{A. Bonnin}
\homepage{Present adress: Kirchhoff-Institut f\"ur Physik, Universit\"at Heidelberg, Im Neuenheimer Feld 227, 69120 Heidelberg, Germany}
\affiliation{LNE-SYRTE, Observatoire de Paris, Universit\'e PSL, CNRS, Sorbonne Universit\'e, \\ 61 Avenue de l'Observatoire, 75014 Paris, France}

\author{C. Solaro}
\homepage{Present adress: Department of Physics and Astronomy, Aarhus University, DK-8000 Aarhus C, Denmark}
\affiliation{LNE-SYRTE, Observatoire de Paris, Universit\'e PSL, CNRS, Sorbonne Universit\'e, \\ 61 Avenue de l'Observatoire, 75014 Paris, France}

\author{X. Alauze}
\affiliation{LNE-SYRTE, Observatoire de Paris, Universit\'e PSL, CNRS, Sorbonne Universit\'e, \\ 61 Avenue de l'Observatoire, 75014 Paris, France}

\author{F. Pereira dos Santos}
\email{franck.pereira@obspm.fr}
\affiliation{LNE-SYRTE, Observatoire de Paris, Universit\'e PSL, CNRS, Sorbonne Universit\'e, \\ 61 Avenue de l'Observatoire, 75014 Paris, France}


\date{\today}

\begin{abstract}
We investigate the collective spin dynamics of a self-rephasing bosonic ensemble of $^{87}$Rb trapped in a 1D vertical optical lattice. We show that the combination of the frequency shifts induced by atomic interactions and inhomogeneous dephasing, together with the spin self-rephasing mechanism leads to the existence of a ``magic density'': \textit{i.e} a singular operating point where the clock transition is first-order insensitive to density fluctuations. This feature is very appealing for improving the stability of quantum sensors based on trapped pseudo-spin-1/2 ensembles. Ramsey spectroscopy of the $|F=1,m_{F}=0\rangle\rightarrow|F=2,m_{F}=0\rangle$ hyperfine transition is in qualitative agreement with a numerical model based on coupled Bloch equations of motion for energy dependent spin vectors.

\end{abstract}


\maketitle


\section{\label{Intro}Introduction}

The coherent manipulation of ultra-cold atoms trapped in optical lattices is of great interest for quantum information processing \cite{Schmidt2005,Lee2013}, quantum simulation \cite{Bloch2012} and high precision measurements. In particular, optical lattices constitute a perfectly relevant framework for the development of highly sensitive and accurate sensors, such as optical lattice clocks \cite{Ushijima2015,Nicholson2015,Schioppo2016,Campbell2017} or inertial sensors \cite{Poli2011,Hilico2015,Ranjit2016}; sensors which subsequently find many outlets for exploring many-body physics \cite{Rey2009,Bishof2011,Martin2013} and various aspects of fundamental physics, such as testing General Relativity \cite{Tarallo2014,Delva2017}, searching for dark matter and drift of fundamental constants \cite{Godun2014,Hees2016} or probing short range forces \cite{Dimopoulos2003,Carusotto2005,Messina2011}.

For many of these applications, high accuracy, sensitivity and stability are required. Accuracy is ensured by isolating, as much as possible, the probed atoms from the external environment and by carefully controlling their interactions with the trapping potentials: by using, \textit{e.g.}, ``magic'' traps \cite{Takamoto2005,Ye2008} for which the two atomic clock states experience exactly the same AC-Stark shift. Sensitivity is greatly improved by simultaneously interrogating a large number of atoms, which, in contrast with single ion experiments, allows for better signal to noise ratio when the measurement is limited by quantum projection noise \cite{Itano1993}. Unfortunately, with a large number of particles, the atom-atom interactions and the resulting density shift \cite{Harber2002,Jannin2015} may also limit the performance of the atomic sensor by adding deleterious bias, noises and drifts. Notably measurements relying on ultracold Fermi atoms, and \textit{a priori} immune to such shifts, were shown to be also impacted by s-wave (and p-wave) collisions when atoms experience an inhomogeneous interrogation field \cite{Campbell2009,Gibble2009,Yu2010}. Different cancellation techniques have then been developed such as tuning the pulses duration in Ramsey \cite{Maineult2012,Hazlett2013} and Rabi \cite{Lee2016} spectroscopy or employing the interaction blockade mechanism \cite{Swallows2011} in a strongly interacting system. As suggested by this latter mechanism, strong interactions can also be beneficial and instigate interesting phenomena such as spin squeezing from many-particle entanglement \cite{Gross2010}, collisional narrowing \cite{Sagi2010} and, of relevance in the work presented here, spin self-rephasing \cite{Deutsch2010,Gibble2010,Buning2011,Solaro2016}.

We explore here the so-called Knudsen regime, where the cumulative effect of the identical spin rotation effect (ISRE) \cite{Lhuillier1982}, which originates from interactions of indistinguishable particles in a non-degenerate gas, counteracts dephasing by keeping atoms synchronized and preserves the coherence of an atomic ensemble for very long times \cite{Deutsch2010,Buning2011}. This collective Spin Self-Rephasing (SSR) mechanism leads to improved sensitivity and stability of the sensor. In that regime, a thorough understanding and modeling of the collective spin dynamics becomes a major issue in order to also ensure a high accuracy.

In this paper, we investigate the collective spin dynamics of a self-rephasing bosonic ensemble of $^{87}$Rb trapped in a shallow 1D vertical optical lattice. We probe the $|F=1,m_{F}=0\rangle\rightarrow|F=2,m_{F}=0\rangle$ hyperfine transition by standard Ramsey spectroscopy. With mean atomic densities $\bar{n}$ around few 10$^{11}$ atoms/cm$^{3}$, the exchange rate of the SSR mechanism $\omega_{\mathrm{ex}}=\frac{4\pi\hbar}{m}a_{12}\bar{n}$ (where $a_{12}$ is the interstate scattering length and $m$ the atomic mass) becomes larger than the typical dephasing rate and results in a non-linear collisional shift. We show that the interplay between this non-linear shift, the differential collisional shift and the frequency chirp induced by inhomogeneities leads to the existence of a singular operating point, for which the clock transition is first-order insensitive to atom density fluctuations. Operating a quantum sensor in this regime, that we name ``magic density'' (with reference to the so-called ``magic wavelength'' \cite{Takamoto2005,Ye2008}), could lead to an increased stability of the measurement. A numerical model, developed previously \cite{Piechon2009}, based on coupled Bloch equations of motion for energy dependent spin vectors qualitatively catches the physics of the problem.

The paper is organized as follows: in section \ref{Section2}, the physical system under study and the experimental set-up are described. In section \ref{Section3}, we study the evolution of the contrast, as well as of the center frequency of the Ramsey fringes, versus Ramsey time and atomic density. We observe and discuss the existence of a magic density and introduce the numerical model used to reproduce our experimental data.

\section{\label{Section2}Physical System and Experimental Set-Up}

In our system, described in details in previous work \cite{Beaufils2011,Pelle2013,Hilico2015}, an ensemble of $^{87}$Rb atoms is trapped in a shallow 1D vertical lattice at wavelength $\lambda_{\mathrm{L}}=532$ nm. The basis can be restricted to two Wannier-Stark ladders, associated to the two clock states $|g\rangle = |5~^{2}S_{1/2},F=1,m_{F}=0\rangle$ and $|e\rangle = |5~^{2}S_{1/2},F=2,m_{F}=0\rangle$. The increment in energy between two consecutive lattice sites is given by the Bloch frequency $\nu_{\mathrm{B}} = 568.5$ Hz. A micro-wave field, resonant with this hyperfine transition $\nu_{\mathrm{HFS}}\sim6.834$ GHz with a Rabi frequency tunable from 1 Hz to few kHz, allows for a standard clock interrogation: \textit{i.e.} two $\pi/2$-pulses separated by the Ramsey time $T_{R}$.

The experimental set-up is described in \cite{Alauze2018} and is dedicated to the measurement of short range forces by trapped atom interferometry \cite{Messina2011}. Initially, few $10^{8}$ atoms of $^{87}$Rb are trapped within 500 ms in a 3D magneto-optical trap (MOT) loaded by a 2D-MOT. After a 100ms stage of compressed MOT, about $10^{7}$ atoms are transferred into an optical dipole trap realized by two crossed beams at $\lambda = 1070$ nm with waists of 30 $\mu$m and 200 $\mu$m and maximum powers of 10 W and 20 W respectively. Evaporative cooling is performed by ramping down the power exponentially to typically 0.1 and 0.25 W within 2 s. The evaporation is stopped just before quantum degeneracy and few $10^{4}$-$10^{5}$ atoms with a temperature in the range 100-500 nK are then adiabatically loaded within 100 ms in the vertical lattice. During evaporation, a sequence of optical pumping, micro-wave pulses and pushing beams polarizes the atomic sample in the state $|F=1,m_{F}=0\rangle$ with more than 99\% efficiency. The vertical lattice is realized by retro-reflecting a 532 nm laser beam with 500 $\mu$m waist and maximal power of 6 W. The transverse confinement is provided by superimposing to the lattice a red detuned progressive wave ($\lambda=1064$ nm, $0.2<P_{\mathrm{IR}}<2$ W, waist of 100 $\mu$m) giving a maximum radial trapping frequency of $\nu_{\mathrm{rad}}\sim45$ Hz. Finally, a cloud of few $10^{4}$ atoms with a transverse size of 40 $\mu$m and a vertical size estimated to few $\mu$m is obtained. The mean density $\bar{n}$ can then be varied from $1\times10^{11}$ atoms/cm$^{3}$ to few 10$^{12}$ atoms/cm$^{3}$ by changing the 2D-MOT loading time.

We explore here the nondegenerate and collisionless Knudsen regime where the trap frequencies are much larger than the lateral collisions rate $\gamma_{c}$. This rate of velocity changing elastic collision is given by $\gamma_{c}\sim4\pi a^{2}v_{\mathrm{T}}\bar{n}$, where $a$ is the scattering length and $v_{\mathrm{T}}=\sqrt{k_{B}T/m}$ the atoms thermal velocity, and corresponds to typically 0.3 s$^{-1}\ll\nu_{\mathrm{rad}}$ in our conditions.

After Ramsey interrogation, the atoms are released from the trap and the populations $N_{g}$ and $N_{e}$ in both hyperfine states are counted via state-selective detection based on fluorescence in horizontal light sheets at the bottom of the vacuum chamber. We then compute the transition probability from $|g\rangle$ to $|e\rangle$: $P = \frac{N_{e}}{N_{g}+N_{e}}$, which depends sinusoidally on the phase difference accumulated during the Ramsey sequence.

\section{\label{Section3}Existence of a Magic Density}

For the SSR mechanism to occur, three conditions are required: (i) the system has to be in the Knudsen regime, (ii) the inhomogeneity of the transition, \textit{i.e.} the inhomogeneous dephasing rate $\Delta_{0}$, must be smaller than the exchange rate $\omega_{\mathrm{ex}}$, (iii) the rate of lateral collisions $\gamma_{c}$ must be smaller than the exchange rate so that the rephasing arises before atoms are scattered in different energy classes.

Following the qualitative two macrospins model introduce in \cite{Deutsch2010}, the SSR mechanism can be readily depicted. The atoms are here divided into two energy classes. After the first $\pi/2$-pulse, these two classes have different transverse spin precession rates because they experience different potentials. In the absence of lateral collision ($\gamma_{c}<\omega_{\mathrm{ex}}$), the atoms do not change class, and dephasing initiates. In the presence of collective and cumulative ISRE, the two classes rephase after an exchange period $T_{\mathrm{ex}} = 2\pi/\omega_{\mathrm{ex}}$ (see \cite{Solaro2016} FIG.2 for a representation on the Bloch sphere). 

This dynamic can also be interpreted by analyzing the system in the singlet-triplet basis of the two-spins states \cite{Gibble2010}: there, dephasing is a precession between the triplet and the singlet states. For bosons, interactions shift the triplet states only (and not the singlet state) due to the symmetry of the wave function. For sufficiently large interaction strengths, the shift of the triplet states gets larger than the dephasing rate, the precession between the triplet and the singlet states is then forbidden and dephasing is inhibited.

To model more quantitatively the spin dynamics we use the model of \cite{Piechon2009,Deutsch2010,Solaro2016}. There, the motion of atoms in the trap is treated semi-classically. In the Knudsen regime ($\gamma_{c}\ll\nu_{\mathrm{rad}}$), the atoms undergo many oscillations in the trap in between two lateral collisions which enables to group the atoms according to their motional energy $E$. The spin dynamics is then described with coupled Bloch equations of motion for energy dependent spin vectors $\mathbf{S}(E,t)$, with $E$ in units of $k_{B}T$:

\begin{equation}\label{Bloch equation of motion}
\begin{array}{lc}
\partial_{t}\mathbf{S}(E,t) \approx & \\
~\left[ \Delta(E)\mathbf{w}+\omega_{\mathrm{ex}}\mathbf{\overline{S}}(t)\right]\times\mathbf{S}(E,t) -~ \gamma_{c}\left[\mathbf{S}(E,t)-\mathbf{\overline{S}}(t)\right] & 
\end{array}
\end{equation}
where $\mathbf{\overline{S}}(t) = \int^{\infty}_{0}dE'g(E')\mathbf{S}(E',t)$ is the energy-average spin ($g(E)$ is the density of states) and $\mathbf{w}$ the vertical unit vector of the Bloch sphere. In our experiment, the dephasing $\Delta(E)$ is dominated by the differential AC-Stark shift of the transverse trapping laser, the mean-field shift and the second order Zeeman shift due to the vertical biais field. It is given by:

\begin{equation}\label{inhomogeneity}
\begin{array}{lcl}
\Delta(E) & = &\left(\delta_{0}^{\mathrm{AC}} + \Delta_{0}^{\mathrm{AC}}E\right) + \Delta_{0}^{\mathrm{MF}}\mathrm{e}^{-E/2} + \delta_{0}^{\mathrm{Z2}} \\
 & = & \delta_{0} + \Delta_{0} E + \mathcal{O}\left(E^{2}\right)
\end{array}
\end{equation}
The homogeneous dephasing $\delta_0$ is responsible for an offset of the hyperfine frequency whose value is the sum of $\delta_{0}^{\mathrm{AC}}/2\pi=-0.7$ Hz, $\delta_{0}^{\mathrm{MF}}/2\pi=\Delta_{0}^{\mathrm{MF}}/2\pi=-0.05\bar{n}$ (with $\bar{n}$ in unit of $10^{11}$ atoms/cm$^{3}$) and $\delta_{0}^{\mathrm{Z2}}/2\pi=4.7$ Hz \footnote{$\delta_{0}^{\mathrm{AC}}$ is given by $(U^{F=2}_{0}-U^{F=1}_{0})/h$, where $U^{F=i}_{0}$ is the trapping potential of the transverse laser experienced by an atom in $F=i$, $i\in\{1,2\}$. $\delta_{0}^{\mathrm{MF}}/2\pi$ is given in Eq.(\ref{Mean-Field shift}) and $\delta_{0}^{\mathrm{Z2}}/2\pi=4.7$ Hz comes from the biais field of 90 mG.}. The linear part $\Delta_{0} = \Delta_{0}^{\mathrm{AC}}-\frac{1}{2}\Delta_{0}^{\mathrm{MF}}$ corresponds to the inhomogeneous dephasing and does not depend on the Zeeman effect as the magnetic field inhomogeneity is negligible among the cloud. In the 2D harmonic approximation, we estimate $\Delta_{0}/2\pi = 0.17 + 0.025\bar{n}$ Hz \footnote{In the harmonic approximation, $\Delta_{0}^{\mathrm{AC}}$ is given by $\frac{k_{B}T}{2\hbar}\times\frac{\delta\alpha}{\alpha}$ \cite{Kuhr2005,Solaro2016}, where $\delta\alpha$ and $\alpha$ are the differential and the total light shift per intensity}. The second term in Eq.(\ref{Bloch equation of motion}), non linear in spin density, represents the exchange mean-field due to ISRE and is responsible for the SSR mechanism. The last term in Eq.(\ref{Bloch equation of motion}) is the effective spin relaxation due to lateral collisions. 

At a given evolution time $T_{R}$ and after an homogeneous $\pi/2$-pulse, as the energy-average spin $\mathbf{\overline{S}}(T_{R})$ stands in the equatorial plane of the Bloch sphere, we can then directly derive the contrast $C(T_{R}) = |\mathbf{\overline{S}}(T_{R})|$ and the frequency measured by Ramsey spectroscopy $\nu(T_{R}) = \arg\left[\mathbf{\overline{S}}(T_{R})\right]/2\pi T_{R}$. In particular in the Knudsen regime, by considering a linear dephasing $\Delta(E)=\Delta_{0}E$ and without exchange interaction, the energy-average spin vector can be expressed as 

\begin{equation}\label{Spin in the plane}
\begin{array}{lcl}
\mathbf{\overline{S}}(T_{R}) & = & S_{x} + iS_{y} \\
 & = & \int^{\infty}_{0}g(E)\exp(i\Delta_{0}E T_{R})dE \\
 & = & \left(1-i\Delta_{0}T_{R}\right)^{-d}
\end{array}
\end{equation}
where $d$ is the dimension of the problem.

\subsubsection{\label{subsection3.1}Study of the Coherence} 

\begin{figure}
\centerline{\includegraphics[width=8.8cm]{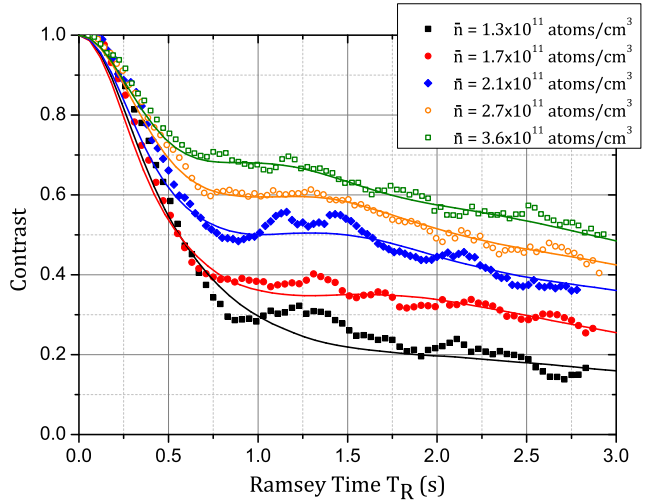}}
\caption
{
(Color online) Contrast versus Ramsey time $T_{R}$ for different mean atomic densities $\bar{n}$. For large $T_{R}$, the contrast increases when $\bar{n}$ increases as a consequence of spin self-rephasing. The solid lines corresponds to fit according to Eq.(\ref{Bloch equation of motion}) with  $\omega_{ex}$, $\gamma_{c}$, $\Delta_{0}^{\mathrm{AC}}$and $\Delta_{0}^{\mathrm{MF}}$ as free fitting parameters.
}
\label{C vs Tr RR0}
\end{figure}

In a first series of experiments, we study the contrast of Ramsey interferometers. The contrast is extracted from the envelop of the Ramsey fringes, which are obtained by scanning both the Ramsey time and the phase of the exciting field. The evolution of the contrast in the lattice for different atomic densities is shown in FIG. \ref{C vs Tr RR0}. It follows the same trend already observed in single well magnetic \cite{Deutsch2010} and optical dipole traps \cite{Buning2011,Solaro2016}: the contrast initially drops because of inhomogeneities and then stabilizes at a value which increases with $\bar{n}$ thanks to SSR. Nevertheless, no clear contrast revivals at the exchange period is observed here.

In order to adjust these results with the model, the following approximations have been made: (i) The radial potential is a 2D harmonic oscillator. (ii) The dynamics in the vertical direction is supposed to be frozen. Indeed, no collision event is likely to change the vertical motion of an atom (\textit{i.e.} Bloch oscillations in that case) as the energy scale of atom-atom interactions is low compared to the Bloch frequency. An ensemble of independent wells is thus considered and the dynamics in each well is modeled using Eq.(\ref{Bloch equation of motion}). (iii) Each of these wells is populated assuming an initial gaussian linear density distribution in $z$. The contrast is then derived by calculating the norm of the sum over all wells of all spins. This explains the absence of sharp revivals in the contrast: as there is a different density in each independent well, this leads to a gaussian distribution of $\omega_{\mathrm{ex}}(z)$ which damps the contrast revivals by averaging over the cloud.

As shown in FIG. \ref{C vs Tr RR0}, our numerical model (solid lines) reproduces well the behaviour of the experimental curves. Nevertheless, as there is no sharp contrast revival, the adjustment of the fit parameters is more delicate than in previous experiments \cite{Deutsch2010,Buning2011,Solaro2016}, as they are not completely independent. The final set of fitting parameters gives us $\omega_{\mathrm{ex}}/2\pi=0.32(2)\times\bar{n}$ Hz and $\gamma_{c}=0.040(4)\times\bar{n}$ s$^{-1}$. These values are respectively smaller by a factor 0.6 and 0.3 than the expected values. This fact, also observed in \cite{Deutsch2010,Buning2011,Solaro2016}, is a consequence of the infinite range approximation taken for these two contributions in Eq.(\ref{Bloch equation of motion}), which overestimates their effects. The inhomogeneities parameters do not exhibit a clear dependency on density and we obtain $\Delta_{0}^{\mathrm{AC}}/2\pi=0.24(5)$ Hz and $\Delta_{0}^{\mathrm{MF}}/2\pi=0.40(8)$ Hz. These values are larger than expected: $\Delta_{0}/2\pi = 0.17 + 0.025\bar{n}$ (with $\bar{n}$ in unit of $10^{11}$ atoms/cm$^{3}$, \textit{cf.} Eq.(\ref{inhomogeneity})). For comparison, by fitting the contrast evolution according to Eq.(\ref{Spin in the plane}) in dimension $d=2$, for the lowest density $\bar{n} = 1.3\times10^{11}$ atoms/cm$^{3}$ (\textit{cf.} black squares in FIG. \ref{C vs Tr RR0}) and for short Ramsey times ($T_{R}<T_{\mathrm{ex}}$), we extract an inhomogeneity of $\Delta_{0}/2\pi = 0.271(2)$ Hz, in better agreement with the previous estimation, showing that experimental parameters are well under control.

As with previous experiments based on single well traps \cite{Deutsch2010,Buning2011,Solaro2016}, this model allows for a good understanding of the coherence of our self-rephasing ensemble of ultra-cold atoms trapped in a 1D lattice. In the following, we further confront this model to our experimental measurement of the center frequency and show that it also reproduces the existence of a magic density.

\subsubsection{\label{subsection3.2}Study of the Frequency}

\begin{figure*}
\centerline{\includegraphics[width=16cm]{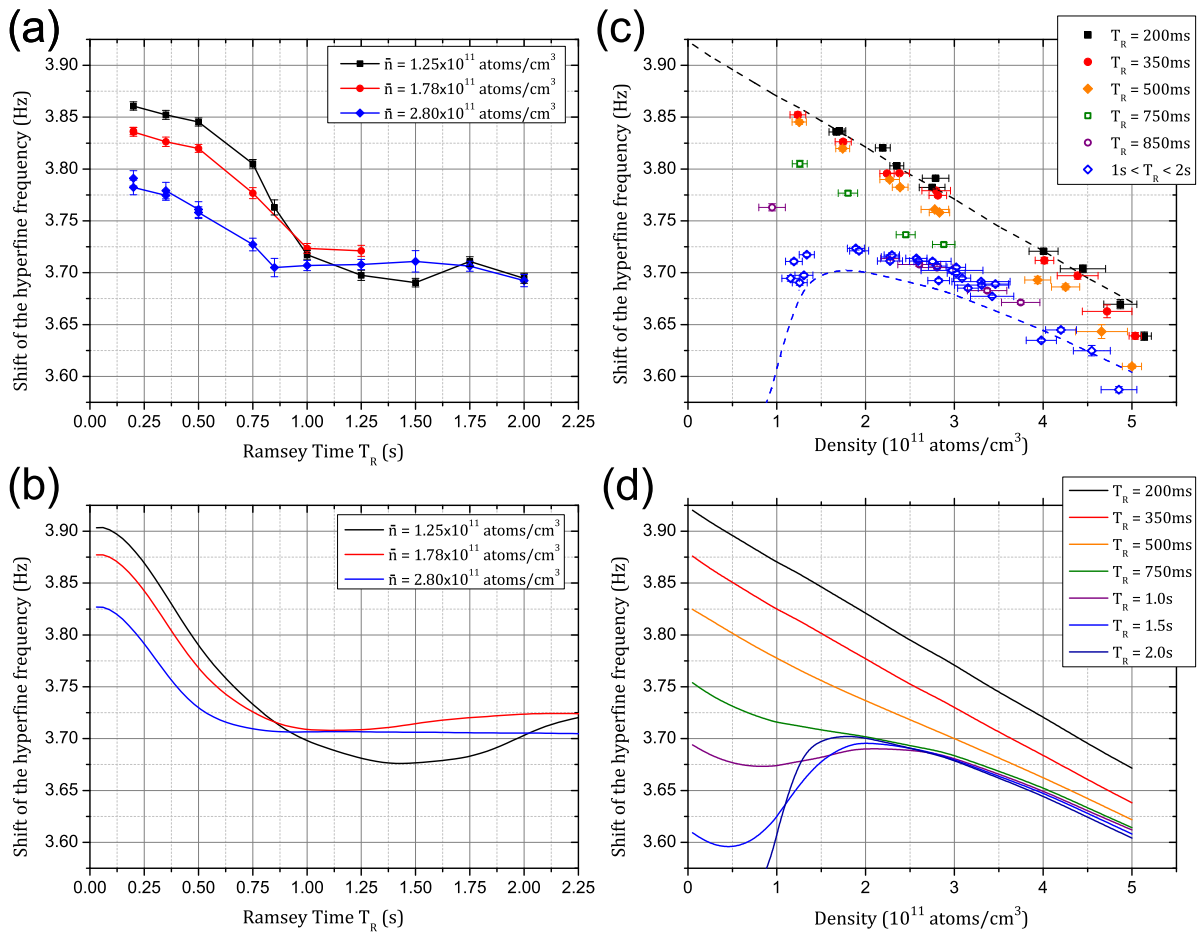}}
\caption
{
\textit{Magic density - } (Color online) - (a) Frequency shift as a function of the Ramsey time for three different values of the atomic density and (b) curves obtained from the numerical model. (c) Frequency shift as a function of the atomic density for different values of the Ramsey time and (d) curves obtained from the numerical model; in (c) the two dashed lines correspond to numerical model for $T_{R} = 200$ ms and $T_{R} = 2$ s. The values of the model's parameters $\omega_{ex}$, $\gamma_{coll}$, $\Delta_{0}^{\mathrm{AC}}$and $\Delta_{0}^{\mathrm{MF}}$ are discussed in the text. The value of the frequency shift at the origin is given by the homogeneous dephasing $\delta_{0}$ (\textit{cf.} Eq.(\ref{inhomogeneity})).
}
\label{Magic density}
\end{figure*}

In our system, three main contributions affect the evolution of the hyperfine frequency during a Ramsey sequence: 

\textit{(i).} Atomic interactions lead to an offset of the clock frequency, linear with the atomic density and given by 

\begin{equation}\label{Mean-Field shift}
\frac{\delta_{0}^{\mathrm{MF}}}{2\pi}=\frac{2\hbar}{m}\left(a_{22}-a_{11}\right)\bar{n}
\end{equation}
where $a_{22}$ and $a_{11}$ are the scattering length for atoms in $|F=2,m_{F}=0\rangle$ and $|F=1,m_{F}=0\rangle$. As $a_{22}<a_{11}$ for $^{87}$Rb, the collisional shift is negative.

\textit{(ii).} The inhomogeneous dephasing gives rise to a chirp of the clock frequency \cite{Kuhr2005}. Using Eq.(\ref{Spin in the plane}) we obtain :

\begin{equation}\label{chirp}
\Delta\nu_{\mathrm{inhom}}(T_{R})=\frac{2}{2\pi T_{R}}\arctan(\Delta_{0}T_{R})
\end{equation}
with a dimension $d=2$.

\textit{(iii).} The collective SSR mechanism counteracts dephasing \cite{Deutsch2010,Buning2011,Solaro2016} and keeps atoms synchronized. Typically, for Ramsey times $T_{R}>T_{\mathrm{ex}}$, the chirp induced by the inhomogeneous dephasing (\textit{cf.} Eq.(\ref{chirp})) is inhibited and the clock frequency remains constant.

From Eq.(\ref{Bloch equation of motion}), the numerical derivation of the clock frequency as a function of the Ramsey time is depicted in FIG.\ref{Magic density}.(b) for three values of the mean atomic density $\bar{n}=\{ 1.25,1.78,2.80\}\times10^{11}$ atoms/cm$^{3}$. The three previous contributions to the clock frequency can be clearly observed: first, the collisional shift creates a negative offset at $T_{R}=0$ which increases linearly with the density. Secondly, the positive inhomogeneity $\Delta_{0}>0$ begets a down-chirp which is, thirdly, counterbalanced by the effect of exchange interactions. The larger the density, the shorter the Ramsey time at which this happens. These simulations are in good agreement with the experimental results shown in FIG.\ref{Magic density}.(a) (the values of the parameters $\omega_{ex}$, $\gamma_{coll}$, $\Delta_{0}^{\mathrm{AC}}$and $\Delta_{0}^{\mathrm{MF}}$ will be discussed later). We can notice the existence of crossing points between the curves for different $\bar{n}$ at given values of $T_{R}$. At these singular operating points, the clock frequency is identical for two different densities and there thus exists in their vicinity a ``magic density'' where the clock frequency is insensitive at first order to density fluctuations. We can also notice that for the highest density $\bar{n}=2.8\times10^{11}$ atoms/cm$^{3}$, where the SSR mechanism is the most efficient, when $T_{R}>T_{\mathrm{ex}}$, the clock frequency remains very stable with the Ramsey time. 

In the following paragraph, we qualitatively discuss a necessary condition for the existence of a magic density. Here, we consider the evolution of the clock frequency as a function of the Ramsey time for two atomic ensembles with different densities $\bar{n}_{1}<\bar{n}_{2}$ and whose inhomogeneities $\Delta_{0}$ are equivalent (see FIG.\ref{Magic density}.(b)). Because of SSR, the chirp induced by the inhomogeneous dephasing will be stopped at shorter Ramsey time for the highest density $\bar{n}_{2}$. In these conditions, if we consider a down-chirp, \textit{i.e.} a positive inhomogeneity $\Delta_{0}>0$, a negative collisional shift (\textit{cf.} Eq.(\ref{Mean-Field shift})) is necessary to obtain a crossing point (and inversely for a down chirp). The sign of the collisional shift $\delta\nu_{\mathrm{MF}}$ and of the inhomogeneous dephasing $\Delta_{0}$ must thus be opposite for the existence of a magic density.

In a second series of experiments, we measure the clock frequency as a function of the atomic density for different Ramsey times (\textit{cf.} FIG.\ref{Magic density}.(c)). There, the atomic density is evaluated by measuring the collisional shift (\textit{cf.} Eq.(\ref{Mean-Field shift})) with Ramsey spectroscopy at short $T_{R}<T_{\mathrm{ex}}$. As suggested by the previous results (\textit{cf.} FIG.\ref{Magic density}.(a) and (b)), for $T_{R}<0.8$ s , there is no crossing points between curves for different $\bar{n}$ and so the collisional shift remains monotonic and no magic density is observed. For $T_{R}>0.8$ s, there exist different crossing points and when $T_{R}>1$ s, all the curves for $1<\bar{n}<3\times10^{11}$ atoms/cm$^{3}$ essentially converge on the same constant frequency shift. In these conditions, a non-monotonic collisional shift is expected and indeed a magic density is clearly observed around $\bar{n}\sim1.8\times10^{11}$ atoms/cm$^{3}$ for $T_{R}>1$ s. On the FIG.\ref{Magic density}.(c) and FIG.\ref{Magic density}.(d), the measurements are qualitatively in good agreement with the numerical model.

Solid lines in FIG.\ref{Magic density}.(c) and (d) are numerical evaluations of Eq.(\ref{Bloch equation of motion}) using only one free fitting parameter $\Delta_{0}^{\mathrm{AC}}$. $\omega_{\mathrm{ex}}$ and $\gamma_{c}$ are extracted from the fit of the contrast study presented in FIG.\ref{C vs Tr RR0}. $\Delta_{0}^{\mathrm{MF}}$ is set to its expected value \footnote{The expected value is $\Delta_{0}^{\mathrm{MF}}\sim \frac{9\sqrt{2}}{2}\frac{4\pi\hbar}{m}\left(a_{22}-a_{11}\right)\bar{n}$, where the factor $\frac{9\sqrt{2}}{2}$ comes from the integration in 2D over $E$ and then on the sum over all the wells in the vertical direction.}. We obtain $\Delta_{0}^{\mathrm{AC}}/2\pi=0.30(2)$ Hz, larger than the expected value but in agreement with the contrast study. For comparison, a fit of the data points of FIG.\ref{Magic density}.(a) by Eq.(\ref{chirp}) for $T_{R}<0.8$ s leads to $\Delta_{0}^{\mathrm{AC}}/2\pi=0.22(2)$ Hz, which again agrees better with the expected value.

\section{\label{Conclusion}Conclusion}

We have shown that the interplay between inhomogeneous dephasing, collisional shift and exchange interaction leads to the existence of a magic density if the sign of inhomogeneous dephasing is opposite to that of the collisional shift. This singular phenomenon, which occurs at an intermediate density regime for which spin self-rephasing starts to counteract the frequency chirp induced by inhomogeneous dephasing, is well reproduced by a numerical model based on coupled Bloch equations of motion for energy dependent spin vectors. The existence of these singular operating points, where the clock transition is insensitive to density fluctuations, is a promising tool to increase the stability and the sensitivity of atomic sensors and especially by tackling one of todays' main limitation in atomic clocks based on trapped ultracold atoms. However, the current relative accuracy on the clock frequency measured in our system (initially not dedicated to perform high precision spectroscopy of internal degrees of freedom) is at a level of few $10^{-12}$. An improved numerical model is consequently necessary if one wants to use the SSR mechanism for a competitive accurate time keeping.

Finally, the magic density can in principle be adjusted by tuning the inhomogeneities. In our experiment, the differential light shift inhomogeneity can not be easily varied (for instance by changing the laser power) without modifying the size of the cloud, and thus the density. Instead, we could use Raman laser beams in a counter-propagating configuration. For a sufficiently low lattice depth, two-photons Raman transitions can coherently and selectively couple given lattice sites thanks to laser induce tunneling \cite{Beaufils2011}. In these conditions, a $\pi/2$-Raman-pulse can be used to shift the atoms in $|F=2\rangle$ by a given number of lattice sites with respect to the atoms in $|F=1\rangle$. This would allow for a fine control of the spatial overlap between atoms in each internal states and subsequently for a fine tuning of the value of $\omega_{\mathrm{ex}}\propto a_{12}\bar{n}$ and thus potentially of the magic density.

\begin{acknowledgments}
We thank F. Combes, J.-N. Fuchs and F. Pi\'echon for insightful discussions and for providing us the code of the numerical model. We also thank P. Rosenbush, W. Maineult and K. Gibble for useful discussions. We acknowledge financial support by the IDEX PSL (ANR-10-IDEX-001-02 PSL) and ANR (ANR-13-BS04-0003-01). A. Bonnin thanks the Labex First-TF for financial support.
\end{acknowledgments}

\bibliography{BiblioCollShift}

\end{document}